\documentclass[conference]{IEEEtran}
\IEEEoverridecommandlockouts

\usepackage{graphicx}
\usepackage{cite}
\usepackage{url}
\usepackage{color}
\usepackage{alltt}
\usepackage{syntax}
\usepackage{newfloat}
\usepackage{mdframed}
\usepackage{multicol}
\usepackage{amsmath}
\usepackage{tikz}
\usepackage{filecontents} 
\usepackage{graphicx}
\usepackage{epstopdf} 
\usepackage{xspace}
\usepackage{enumitem}
\usepackage{svg}
\usepackage{lscape}
\usepackage{longtable}
\usepackage{supertabular}

\usepackage[ruled,linesnumbered,noend]{algorithm2e}
\SetKw{Continue}{continue}
\SetKwInput{KwInput}{Input}
\SetKwInput{KwOutput}{Output}
\SetKwProg{Fn}{Function}{}{end}

\SetCommentSty{mycommfont}

\usepackage{adjustbox}
\usepackage{wrapfig}
\usepackage{graphbox}
\usepackage{rotating}
\usepackage{pdfpages}
\usepackage{mdwlist}
\usepackage{multirow}
\usepackage[T1]{fontenc}

\usepackage{subcaption}
\usepackage[labelfont=bf,skip=0pt]{caption}
\usepackage{hyperref} 
\hypersetup{
	citebordercolor={0 1 0},
	citecolor=blue,
	colorlinks=false,
	filebordercolor={0 .5 .5},
	filecolor=green,
	linkbordercolor={1 0 0},
	linkcolor=blue,
	menubordercolor={1 0 0},
	pageanchor=true,
	pagebackref=true,
	pdfborder={0 0 1},
	pdfpagelabels=true,
	pdftex,
	plainpages=false,
	urlbordercolor={0 1 1},
	urlcolor=blue,
}

\usepackage[capitalize,nameinlink]{cleveref}
\hypersetup{%
	bookmarksnumbered, bookmarksopen=true, bookmarksopenlevel=1,%
}
\crefname{figure}{Figure}{Figures}
\crefname{listing}{Query}{Queries}
\crefname{section}{Section}{Sections}
\crefname{table}{Table}{Tables}
\crefname{BNF}{Grammar}{Grammars}
\crefname{algorithm}{Algorithm}{Algorithms}
\crefname{equation}{Equation}{Equations}

\usepackage{listings}
\definecolor{mygreen}{rgb}{0,0.6,0}
\definecolor{mygray}{rgb}{0.5,0.5,0.5}

\newcommand{\incode}[1]{\lstinline{#1}}

\newcommand{\distance}{5pt}
\DeclareFloatingEnvironment[
fileext   = logr,
listname  = {List of Grammars},
name      = Grammar,
placement = !htbp
]{BNF}

\newcommand{\msim}{\raise.17ex\hbox{$\scriptstyle\sim$}}

\newcommand{\myparatight}[1]{\smallskip\noindent{\bf {#1}.}}

\newcommand{\eat}[1]{}
\newcommand{\eg}{e.g.,\xspace}
\newcommand{\ie}{i.e.,\xspace}

\newcommand{\tool}{\textsc{ThreatRaptor}\xspace}
\newcommand{\lang}{TBQL\xspace}
\newcommand{\cti}{OSCTI\xspace}

\def\BibTeX{{\rm B\kern-.05em{\sc i\kern-.025em b}\kern-.08em
    T\kern-.1667em\lower.7ex\hbox{E}\kern-.125emX}}

\begin{document}

\title{A System for Efficiently Hunting for Cyber Threats in Computer Systems Using Threat Intelligence}

\author{
\IEEEauthorblockN{Peng Gao\IEEEauthorrefmark{1},
Fei Shao\IEEEauthorrefmark{2},
Xiaoyuan Liu\IEEEauthorrefmark{1},
Xusheng Xiao\IEEEauthorrefmark{2},
Haoyuan Liu\IEEEauthorrefmark{1},
Zheng Qin\IEEEauthorrefmark{3},
Fengyuan Xu\IEEEauthorrefmark{3}\\
Prateek Mittal\IEEEauthorrefmark{4},
Sanjeev R. Kulkarni\IEEEauthorrefmark{4},
Dawn Song\IEEEauthorrefmark{1}}
\IEEEauthorblockA{\IEEEauthorrefmark{1}University of California, Berkeley    \IEEEauthorrefmark{2}Case Western Reserve University \\   \IEEEauthorrefmark{3}National Key Lab for Novel Software Technology, Nanjing University
\IEEEauthorrefmark{4}Princeton University \\
\IEEEauthorrefmark{1}\{penggao,xiaoyuanliu,hy.liu,dawnsong\}@berkeley.edu    \IEEEauthorrefmark{2}\{fxs128,xusheng.xiao\}@case.edu   \\ \IEEEauthorrefmark{3}\{qinzheng,fengyuan.xu\}@nju.edu.cn
\IEEEauthorrefmark{4}\{pmittal,kulkarni\}@princeton.edu
}
}

\maketitle

\begin{abstract}

Log-based cyber threat hunting has emerged as an important solution to counter sophisticated cyber attacks.
However, existing approaches require non-trivial efforts of manual query construction and have overlooked the rich external knowledge about threat behaviors provided by open-source Cyber Threat Intelligence (\cti).
To bridge the gap, we build \tool, a system that facilitates cyber threat hunting in computer systems using \cti.
Built upon mature system auditing frameworks, \tool provides (1) an unsupervised, light-weight, and accurate NLP pipeline that extracts structured threat behaviors from unstructured \cti text,
(2) a concise and expressive domain-specific query language, \lang, to hunt for malicious system activities,
(3) a query synthesis mechanism that automatically synthesizes a \lang query from the extracted threat behaviors,
and 
(4) an efficient query execution engine to search the big system audit logging data.

\end{abstract}

\section{Introduction}
\label{sec:intro}

\begin{figure*}[t]
    \centering
    \includegraphics[width=0.7\linewidth]{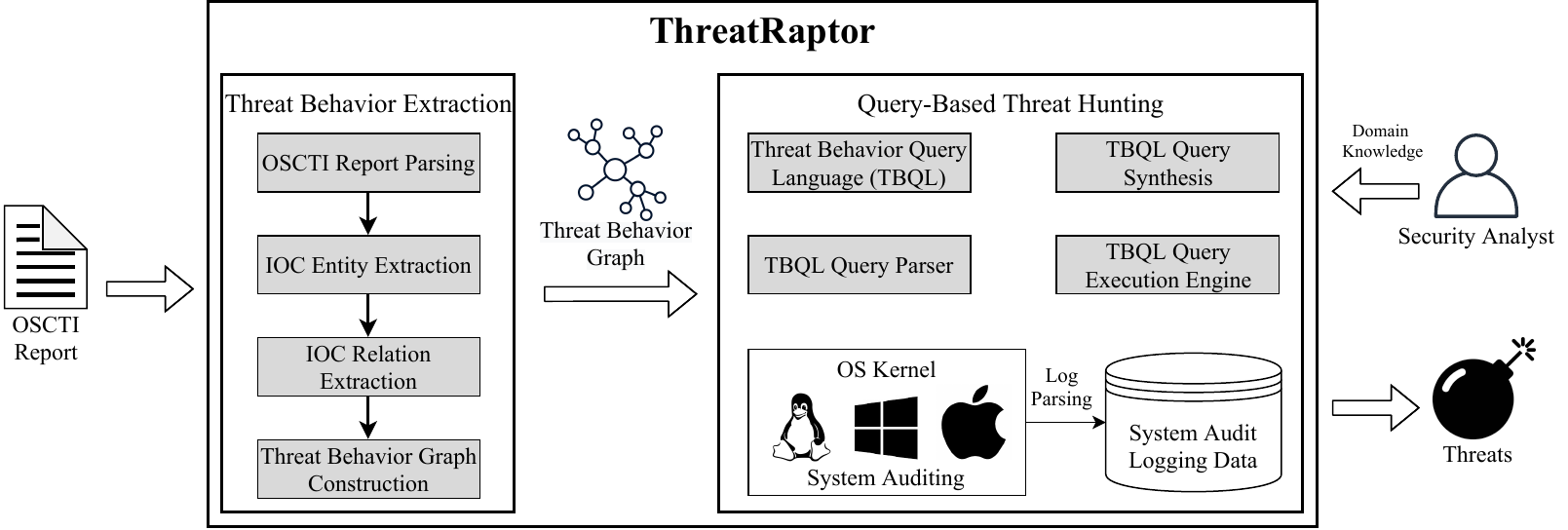}
    \caption{The architecture of \tool}
    \label{fig:arch}
\end{figure*}

Recent cyber attacks have plagued many high-profile
businesses~\cite{equifax}.
These attacks often exploit multiple types of vulnerabilities to infiltrate into target systems in multiple stages.
To counter these attacks, \emph{ubiquitous system auditing} has emerged as an important approach for monitoring system activities.
System auditing collects system-level auditing events about system calls from OS kernel as system audit logs.
The collected audit logging data further enables approaches to hunt for cyber threats via query processing~\cite{gao2018aiql,gao2018saql,gao2019query,gao2020querying}.

Cyber threat hunting in enterprises is the process of proactively and iteratively searching for malicious 
activities in various types of logs, which is critical to early-stage detection.
Despite numerous efforts~\cite{gao2018aiql,splunk-spl},
existing approaches, however, require non-trivial efforts of manual query construction and have overlooked the rich external threat knowledge
provided by open-source Cyber Threat Intelligence (\cti).
Hence, the threat hunting process is labor-intensive and error-prone.

\cti~\cite{os-cti} is a form of evidence-based knowledge and has received growing attention from the community.
Commonly, knowledge about threats is presented in a vast number of publicly available \cti sources.
Structured \cti feeds~\cite{stix} have primarily focused on Indicators of Compromise (IOCs), such as malicious file/process names and IP addresses.
Though useful in capturing fragmented views of threats, these disconnected IOCs lack the capability to uncover the complete threat scenario as to how the threat unfolds into multiple steps.
In contrast, unstructured \cti reports~\cite{securelist} contain more comprehensive threat knowledge.
For example, descriptive relationships between IOCs contain knowledge about multi-step threat behaviors
(\eg ``read'' relationship between two IOCs ``/bin/tar'' and ``/etc/passwd'' in \cref{fig:demo}), which is critical to uncovering the complete threat scenario.
Unfortunately, none of the existing approaches provide an automated way to harvest such knowledge and use it for threat hunting.

There are two major challenges for building a system that (1) extracts knowledge about threat behaviors (IOCs and IOC relationships) from unstructured \cti reports, and
(2) uses the 
knowledge for threat hunting.
First, accurately extracting threat knowledge from natural-language \cti text is not trivial.
This is due to the presence of massive nuances particular to the security context, such as special characters (\eg dots, underscores) in IOCs.
These nuances limit the performance of most NLP modules (\eg tokenization).
Second, system auditing often produces a huge amount of daily logs (0.5 GB $\sim$ 1 GB for 1 enterprise host~\cite{reduction}), and hence threat hunting is a procedure of ``finding a needle in a haystack''. 
Such a big amount of log data poses challenges for solutions to store and query the data efficiently to hunt for malicious 
activities.

To address these challenges,
we build \tool, a system that facilitates threat hunting in computer systems using \cti.
\tool ($\sim25$K LOC) was built upon mature system auditing frameworks
for system audit logging data collection and databases
for data storage.
Particularly, \tool has four novel designs:
(1) An unsupervised, light-weight, and accurate NLP pipeline for extracting threat behaviors (IOCs and IOC relations) from \cti texts.
The pipeline employs a series of techniques (\eg IOC protection, dependency parsing-based IOC relation extraction) to handle nuances and perform accurate extraction.
The extracted IOCs and IOC relations form a \emph{threat behavior graph}, which is amenable to automated processing;
(2) A concise and expressive domain-specific query language, \emph{Threat Behavior Query Language (\lang)}, for querying system audit logging data stored in different database backends.
\lang is a declarative query language that uniquely integrates a series of 
primitives for threat hunting in computer systems (\eg system entities, system events, event path patterns, various types of filters);
(3) A query synthesis mechanism for automatically synthesizing a \lang query from the threat behavior graph;
(4) A query execution engine for efficiently executing \lang queries.
To the best of our knowledge, \tool is \emph{the first system that bridges \cti with system auditing for threat hunting}.
For more details, please refer to our full-length paper~\cite{gao2021enabling}.

\textbf{Demo video:} \url{https://youtu.be/SrcTDQwRF_M}

\section{The \tool Architecture}
\label{sec:architecture}

\begin{figure*}[t]
\centering
    \includegraphics[width=\linewidth]{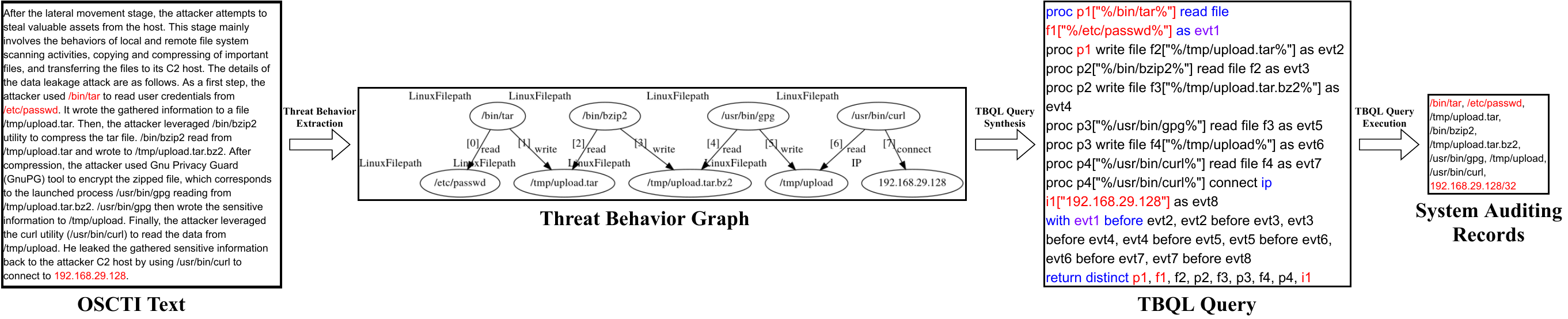}
    \caption{An example data leakage attack case demonstrating the whole processing pipeline of \tool}
    \label{fig:demo}
\end{figure*}

\cref{fig:arch} shows the architecture of \tool.
Given an input \cti report, \tool extracts IOCs (\eg file names, file paths, IPs) and IOC relations, constructs a threat behavior graph, synthesizes a \lang query, and executes the synthesized query to retrieve the matched system auditing records.
\cref{fig:demo} shows an example data leakage attack case demonstrating the whole pipeline.

\subsection{Data Collection}
\label{subsec:collection}

System audit logging data records the interactions among system entities as system events. 
Following the established convention~\cite{gao2018aiql,reduction}, 
we consider system entities as \emph{files}, \emph{processes}, and \emph{network connections}.
We consider a system event as the interaction between two system entities represented as $\langle$subject, operation, object$\rangle$.
Subjects are processes originating from software applications (\eg Chrome), and objects can be files, processes, and network connections.
We categorize system events into three types according to the types of their object entities: \emph{file events}, \emph{process events}, and \emph{network events}.

\tool leverages a mature system auditing framework, Sysdig, to collect system audit logs from a host.
\tool then parses the collected logs into system entities and system events, and extracts critical attributes.
Representative entity attributes are: file name, process executable name, src/dst IP, src/dst port.
Representative event attributes are: sbj/obj entity ID, operation, start/end time.

\subsection{Data Storage}
\label{subsec:storage}

\tool leverages a relational database, PostgreSQL, and a graph database, Neo4j, for its storage component.
Relational databases come with mature indexing mechanisms and are scalable to massive data, which are suitable for queries that involve many joins and constraints.
Graph databases represent data as nodes and edges, 
which are suitable for queries that involve graph pattern search.
For PostgreSQL, \tool stores system entities and system events in tables.
For Neo4j, \tool stores system entities as nodes and system events as edges.
Indexes are created on key attributes 
to speed up the search.
Furthermore, to reduce the data size, \tool leverages the  Causality Preserved Reduction technique~\cite{reduction} to merge excessive events between the same pair of entities.

\subsection{Threat Behavior Extraction}
\label{subsec:extraction}

\tool employs a specialized NLP pipeline (built upon spaCy) to accurately extract IOCs and IOC relations and construct a threat behavior graph (\cref{alg:threat-behavior-extraction}).

(1) 
\emph{Block Segmentation (Line 3) and Sentence Segmentation (Line 6):}
We segment an input \cti article into natural blocks.
We then segment a block into sentences.

(2) 
\emph{IOC Recognition and IOC Protection (Line 5):}
We construct a set of regex rules to recognize various types of IOCs (\eg file name, file path, IP).
Furthermore, we protect the security context by replacing the IOCs with a dummy word (\ie word ``something'').
This makes the NLP modules designed for processing general text work well for \cti text.

(3)
\emph{Dependency Parsing (Line 7):}
We construct a dependency tree for each sentence.
We then replace the dummy word with the original IOCs in the trees.

(4)
\emph{Tree Annotation (Line 9):}
We annotate nodes in the dependency trees whose associated tokens are useful for coreference resolution and relation extraction tasks (\eg IOCs, candidate IOC relation verbs, pronouns).

(5)
\emph{Tree Simplification (Line 10):}
We simplify the annotated trees by removing paths without IOC nodes down to the leaves.

(6)
\emph{Coreference Resolution (Line 13):}
Across all trees of all sentences within a block, we resolve the coreference nodes for the same IOC by checking their POS tags and dependencies,
and create connections between the nodes in the trees.

(7)
\emph{IOC Scan and Merge (Line 15):}
We scan all IOCs in the 
trees of all blocks, and merge similar ones based on both the character-level overlap and the word vector similarities.

(8)
\emph{IOC Relation Extraction (Line 17):}
For each dependency tree, we enumerate all pairs of IOCs nodes. 
Then, for each pair, we check whether they satisfy the subject-object relation by considering their dependency types in the tree.
In particular, we consider three parts of their dependency path: one common path from the root to the LCA (Lowest Common Ancestor); two individual paths from the LCA to each of the nodes, and construct a set of dependency type rules to do the checking.
Next, for the pair that passes the checking, we extract its relation verb by first scanning all the annotated candidate verbs in the aforementioned three parts of dependency path, and then selecting the one that is closest to the object IOC node.
The candidate IOC node pair and the selected verb (after lemmatization) then form the final IOC entity-relation triplet.

(10)
\emph{Threat Behavior Graph Construction (Line 19):}
We iterate over all IOC entity-relation triplets sorted by the occurrence offset of the relation verb in \cti text, and construct a threat behavior graph.
Each edge in the graph is associated with a sequence number, indicating the step order.

\begin{algorithm}[t]
\footnotesize

\SetAlgoLined
\SetKwData{Doc}{document}\SetKwData{Blk}{block}\SetKwData{RR}{replacementRecord}\SetKwData{Sent}{sentence}\SetKwData{Tree}{tree}\SetKwData{Trees}{trees}\SetKwData{Treess}{all_block_trees}\SetKwData{Bg}{graph}\SetKwData{Ent}{entities}\SetKwData{AEnt}{all_iocs}\SetKwData{Rel}{ioc_rels}\SetKwData{ARel}{all_ioc_rels}
\SetKwFunction{BSeg}{SegmentBlock}\SetKwFunction{NEP}{ProtectIoc}\SetKwFunction{SSeg}{SegmentSentence}\SetKwFunction{DP}{ParseDependency}\SetKwFunction{RNEP}{RemoveIocProtection}\SetKwFunction{TL}{AnnotateTree}\SetKwFunction{Coref}{ResolveCoref}\SetKwFunction{LBSP}{SimplifyTree}\SetKwFunction{BGG}{ConstructBehaviorGraph}\SetKwFunction{SRE}{ScanMergeIoc}\SetKwFunction{AE}{MergeSimilarEntities}\SetKwFunction{ME}{ExtractIocRelation}\SetKwFunction{GG}{ConstructGraph}
\SetKwInOut{Input}{Input}
\SetKwInOut{Output}{Output}
\BlankLine
\Input{OSCTI Text: \Doc}
\Output{Threat Behavior Graph: \Bg}
\BlankLine

Initialize \Treess\;
Initialize \ARel\;
\For{\Blk in \BSeg{\Doc}}{
    Initialize \Trees\;
    \Blk $\leftarrow$ \NEP{\Blk}\;
    \For{\Sent in \SSeg{\Blk}}{
        \Tree $\leftarrow$ \DP{\Sent}\;
        \Tree $\leftarrow$ \RNEP{\Tree}\;
        \Tree $\leftarrow$ \TL{\Tree}\;
        \Tree $\leftarrow$ \LBSP{\Tree}\;
        Add \Tree to \Trees\;
    }
    \For{\Tree in \Trees}{
        \Tree $\leftarrow$ \Coref{\Tree, \Trees}\;
    }
    Add all \Tree in \Trees to \Treess\; 
}

\AEnt $\leftarrow$ \SRE{\Treess}\;
\For{\Tree in \Trees}{
    \Rel $\leftarrow$ \ME{\Tree, \Trees, \AEnt}\;
    Add \Rel to \ARel\;
}
\Bg $\leftarrow$ \GG{\AEnt, \ARel}\;

\caption{Threat Behavior Extraction Pipeline}
\label{alg:threat-behavior-extraction}
\end{algorithm}

\subsection{Threat Behavior Query Language (\lang)}
\label{subsec:language}

\tool provides a domain-specific language, \lang (built upon ANTLR 4), to query system audit logging data.
Compared to general-purpose query languages (\eg SQL, Cypher) that are low-level and verbose, \lang treats system entities and
events as first-class citizens and provides 
primitives to easily specify multi-step system activities.

The basic \emph{event pattern syntax} of \lang specifies one or more system event patterns in the format of $\langle$subject, operation, object$\rangle$,
with optional filters on the temporal and attribute relationships between event patterns.
System entities have explicit types and identifiers, with optional filters on the entity attributes.
Operators (\eg logical, comparison) are supported in event operations and attribute filters to form complex expressions.
Optional time windows can be specified for event patterns to constrain the search.

\cref{fig:demo} shows an example synthesized \lang query in this syntax.
Eight event patterns are declared (\incode{evt1} - \incode{evt8}), with entity types, identifiers, and attribute filters.
The {\tt with} clause specifies the temporal orders of events (\ie temporal relationships).
Besides, several syntactic sugars are adopted to make the query concise:
(1) default attribute names are omitted in the event patterns and the {\tt return} clause, which will be inferred during query execution.
We select the most commonly used attributes in security analysis as default attributes: ``name'' for files, ``exename'' for processes, and ``dstip'' for network connections.
For example, \incode{proc p1["\%/bin/tar\%"]} will be inferred as \incode{proc p1[exename = "\%/bin/tar\%"]}
and \incode{return p1} will be inferred as \incode{return p1.exename};
(2) an entity ID is used in multiple event patterns, which means that the referred entities are the same.
For example, \incode{p1} is used in both \incode{evt1} and \incode{evt2}, which is equivalent to an attribute relationship between the two event patterns, \ie
\incode{evt1.srcid = evt2.srcid}.

Besides the basic syntax, \tool provides an advanced syntax that specifies 
\emph{variable-length paths of system event patterns}.
This syntax is particularly useful when doing query synthesis: in some cases, an edge in the threat behavior graph may correspond to a path of system events in system audit logging data.
This happens often when intermediate processes are forked to chain system events, but are omitted in the \cti text by the human writer. 
For example, \incode{proc p \~>[read] file f} specifies a path of arbitrary length from a process entity 
\incode{p} to a file entity \incode{f}. 
The operation type of the final hop
is \incode{read}.
\incode{proc p \~>(2~4)[read] file f} furthers specifies the minimum and maximum lengths of the path.
More language features are illustrated in our demo video.


\subsection{\lang Query Synthesis}
\label{subsec:synthesis}

\tool provides a query synthesis mechanism that automatically synthesizes a \lang query from the threat behavior graph. 
The synthesis starts with a screening to filter out nodes (and connected edges) in the threat behavior graph whose associated IOC types are not currently captured by the system auditing component.
Then, for each remaining edge, \tool maps its associated IOC relation to the \lang operation type using a set of rules (\eg the ``download'' relation between two ``Filepath'' IOCs will be mapped to the ``write'' operation in \lang, indicating a process writes data to a file).
Next, \tool synthesizes the subject/object entity from the source/sink node,
and synthesizes an event pattern by connecting the entities with the operation.
\tool then synthesizes the temporal relationships of the event patterns in the {\tt with} clause based on the sequence numbers of the corresponding edges.
Finally, \tool synthesizes the {\tt return} clause by appending all entity IDs. 
In addition to the default synthesis plan, \tool supports user-defined plans to synthesize other patterns (\eg path patterns) and attributes (\eg time window).

\subsection{\lang Query Execution}
\label{subsec:execution}

To execute a \lang query with multiple patterns, \tool compiles each pattern into a semantically equivalent SQL or Cypher data query, and schedules the execution of these data queries in different database backends.
Specifically, for an event pattern, \tool compiles it into a SQL data query which joins entity tables with event table.
For a variable-length event path pattern, since it is difficult to perform graph pattern search using SQL, \tool compiles it into a Cypher data query by leveraging Cypher's path pattern syntax.

For each pattern, \tool computes a pruning score 
by counting the number of constraints declared; a pattern with more constraints has a higher score.
For a variable-length event path pattern, \tool additionally considers the path length;
a pattern with a smaller maximum path length has a higher score.
Then, when scheduling the execution of the data queries, \tool considers both the pruning scores and the pattern dependencies: 
if two patterns are connected by the same system entity, 
\tool will first execute the data query whose associated pattern has a higher pruning score, and then use the execution results to constrain the execution of the other data query (by adding filters).
This way, complex \lang queries can be efficiently executed in different database backends seamlessly.

\section{Demonstration Outline}
\label{sec:demo}

\begin{figure}[t]
    \centering
    \includegraphics[width=0.73\linewidth]{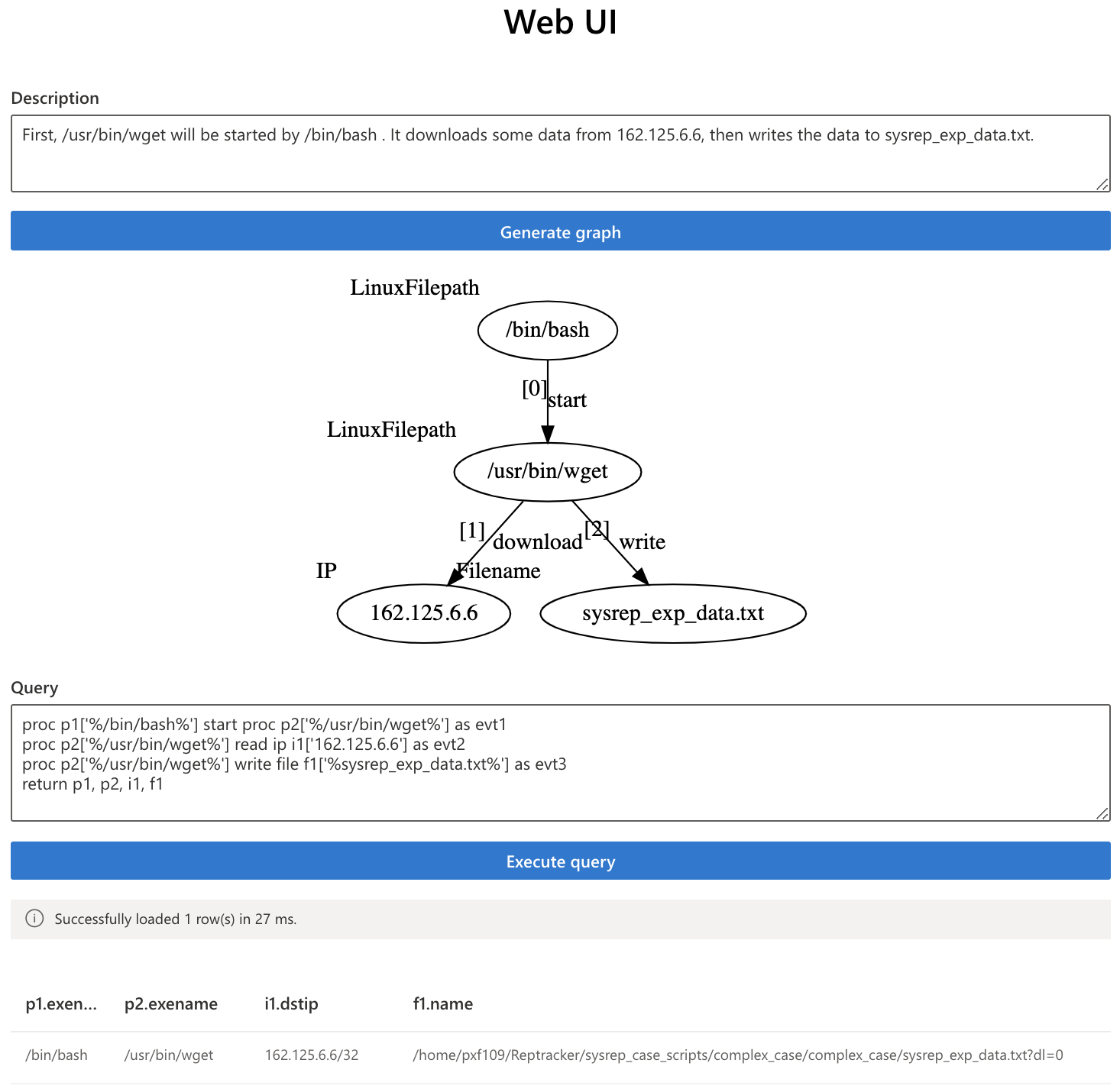}
    \caption{The web UI of \tool}
    \label{fig:ui-demo}
\end{figure}

We deployed \tool on a server and built a web UI (\cref{fig:ui-demo}).
In our demo, we aim to show the complete usage scenario of \tool. 
We perform two multi-step intrusive attacks on the deployed server and construct attack descriptions according to the way the attacks were performed.
Constructed based on CVE~\cite{cve}, these two attacks exploit system vulnerabilities and exfiltrate sensitive data:

\begin{itemize}[noitemsep, topsep=1pt, partopsep=1pt, listparindent=\parindent, leftmargin=*]
    \item \emph{Password Cracking After Shellshock Penetration:} The attacker penetrates into the victim host (\ie the deployed server) by exploiting the Shellshock vulnerability.
    After the penetration, the attacker first connects to cloud services (Dropbox) and downloads an image where C2 (Command and Control) server's IP address is encoded in the EXIF metadata. 
    Based on the IP address, the attacker downloads a password cracker from the C2 server to the victim host.
    The attacker then runs the password cracker against password shadow files to extract clear text.

    \item \emph{Data Leakage After Shellshock Penetration:} 
    The attacker attempts to steal all the valuable assets from the victim host. 
    The attacker scans the file system, scrapes files into a single compressed file, and transfers it back to the C2 server.
\end{itemize}

When we perform the attacks, the server continues to resume its routine tasks to emulate the real-world deployment,
where benign system activities and malicious system activities co-exist.
We use \tool to hunt for malicious system activities by feeding the attack descriptions to the system, which in turn synthesizes \lang queries and executes the synthesized queries over the collected data.
The audience will have the option to conduct the attacks and perform threat hunting through \tool's web UI.

\section{Conclusion}
We have presented \tool, a novel system that 
facilitates
threat hunting in computer systems using \cti. 

\myparatight{Acknowledgement}
This work was supported in part by DARPA N66001-15-C-4066 and the CLTC (Center for Long-Term Cybersecurity).

\bibliographystyle{IEEEtran}
\bibliography{refs}

\end{document}